\documentclass{desyproc}

\begin{document}
\title{Constraints on pseudoscalar-photon interaction from CMB polarization observation}

\author{{\slshape Wei-Tou Ni$^1$$^,$$^2$}\\[1ex]
$^1$Center for Gravitation and Cosmology, Department of Physics, National Tsing Hua University,
 Hsinchu, Taiwan, 30013 ROC  \ \ Email: weitou@gmail.com\\
$^2$Purple Mountain Observatory, Chinese Academy of Sciences,
Nanjing 210008, China}

\contribID{lindner\_axel}

\desyproc{DESY-PROC-2009-05}
\acronym{Patras 2009} 
\doi  
\maketitle

\begin{abstract}
Effective pseudoscalar-photon interaction(s) would induce a rotation
of linear polarization of electromagnetic wave propagating with
cosmological distance in various cosmological models.
Pseudoscalar-photon interaction is proportional to the gradient of
the pseudoscalar field. From phenomenological point of view, this
gradient could be neutrino number asymmetry, other density current,
or a constant vector. In these situations, Lorentz invariance or CPT
may effectively be violated. CMB polarization observations are
superb tests of these models and have the potential to discover new
fundamental physics. In this paper, we review the constraints on
pseudoscalar-photon interaction from CMB polarization observation.
\end{abstract}

\section{Introduction}

In 1973, we studied the relationship of Galilio Equivalence Principe
(WEP I) and Einstein Equivalence Principle in a framework (the
$\mathrm{\mathrm{\chi}}$-$\mathrm{g}$ framework) of electromagnetism
and charged particles, we found the following theory with
(gravitational) interaction Lagrangian density

\begin{eqnarray}
L_\mathrm{int} =
-(\frac{1}{16\pi})(-g)^{1/2}[\frac{1}{2}g^{ik}g^{jl}-\frac{1}{2}g^{il}g^{kj}+\phi\varepsilon^{ijkl}]F_{ij}F_{kl}
- A_kj^k(-g)^{1/2} - \Sigma_I \frac{ds_I}{dt}\delta(x - x_I),
\label{eqn1}
\end{eqnarray}
as an example which obeys WEP I, but not EEP \cite{ref1,ref2,ref3}.
The nonmetric part of this theory is

\begin{eqnarray}
L^{(\emph{NM})}_{\ \ \ \ \ \mathrm{int}} =
-(\frac{1}{16\pi})(-g)^{1/2}\phi\varepsilon^{ijkl}F_{ij}F_{kl} =
-(\frac{1}{4\pi})(-g)^{1/2}\phi_{,i} \varepsilon^{ijkl}A_jA_{k,l} \
(\mathrm{mod \ div}), \label{eqn2}
\end{eqnarray}
where `mod div' means that the two Lagrangian densities are related
by partial integration in the action integral. The Maxwell equations
\cite{ref1,ref3} are

\begin{eqnarray}
F^{ik}_{\ \ \mid k} + \varepsilon^{ikml}F_{km}\varphi_{,l}=-4\pi
j^i, \label{eqn3}
\end{eqnarray}
where the derivation $|$ is with respect to the Christoffel
connection. The Lorentz force law is the same as in metric theories
of gravity or general relativity. Gauge invariance and charge
conservation are guaranteed. The Maxwell equations are also
conformally invariant.

The rightest term in equation (2) is reminiscent of Chern-Simons
\cite{ref4} term $e^{\alpha\beta\gamma}A_\alpha F_\beta\gamma$.
There are two differences: (i) Chern-Simons term is in 3 dimensional
space; (ii) Chern-Simons term is a total divergence.

A term similar to the one in equation (2) (axion-gluon interaction)
occurs in QCD in an effort to solve the strong CP problem (Peccei
and Quinn \cite{ref5}, Weinberg \cite{ref6}, Wilczek \cite{ref7}).
Carroll, Field and Jackiw \cite{ref8} proposed a modification of
electrodynamics with an additional $e^{ijkl} V_i A_j F_{kl}$ term
with {\it V}$_i$ a constant vector. This term is a special case of
the term $e^{ijkl} \varphi F_{ij} F_{kl}$ (mod div) with
$\varphi_{,i}$ = $-\frac{1}{2}V_i$   .

 Various terms in the Lagrangians discussed in this section are
listed in Table \ref{table1}. Empirical tests of the
pseudoscalar-photon interaction (2) from CMB polarization
observation will be discussed in section 2. Section 3 will present
an outlook.

\begin{table}[ph]
\centerline{\begin{tabular}{c c c c}
\hline \\[-10pt] Term & Dimension &
Reference & Meaning \\[2pt]
\hline \\[2pt] $e^{\alpha\beta\gamma}A_\alpha F_{\beta\gamma}$ & 3 &
Chern-Simons (1974\cite{ref4}) & Integrand for topological  \\[-2pt]
& & &invariant\\[5pt]
$e^{ijkl}\varphi F_{ij}F_{kl}$ & 4 & Ni (1973[1],
1974\cite{ref2},1977\cite{ref3})
 & Pseudoscalar-photon\\[-2pt]
  & & &coupling \\[3pt]
$e^{ijkl}\varphi F^{QCD}_{\ \ \ \ \ \ ij}F^{QCD}_{\ \ \ \ \ \ kl}$ & 4 & Peccei-Quinn (1977\cite{ref5}) & Pseudoscalar-gluon  \\[-2pt]
& &Weinberg (1978\cite{ref6})& coupling \\[-13pt]
\\& &Wilczek (1978\cite{ref7}) \\[5pt]
 $e^{ijkl}V_iA_j F_{kl}$ & 4 & Carroll-Field-Jackiw  &External constant \\[-2pt]
& &(1990\cite{ref8})& vector coupling \\[10pt]
\hline
\end{tabular}}
\caption{Various terms in the Lagrangian and their meaning.}
\end{table} \label{table1}

\section{Constraints from CMB polarization observation}

Pseudoscalar-photon interaction induces polarization rotation in
electromagnetic propagation. From (3), for the right circularly
polarized electromagnetic wave, the propagation from a point P$_1$
to another point P$_2$ adds a phase of $\alpha =
\varphi(\mathrm{P}_2) - \varphi(\mathrm{P}_1)$ to the wave; for left
circularly polarized light, the added phase will be opposite in sign
\cite{ref1}. Linearly polarized electromagnetic wave is a
superposition of circularly polarized waves. Its polarization vector
will then rotate by an angle $\alpha$. When the propagation distance
is over a large part of our observed universe, we call this
phenomenon cosmic polarization rotation \cite{ref9,ref10}.

Since the first successful polarization observation of the
cosmological microwave background (CMB) in 2002 by DASI \cite{ref11}
(Degree Angular Scale Interferometer), there have been a number of
observations [12-16] with better precision. These observations set
up limits on the electromagnetic polarization rotation due to
effective pseudoscalar-photon interaction.

In the CMB polarization observations, there are variations and
fluctuations. The variations and fluctuations due to scalar-modified
propagation can be expressed as $\delta\varphi(2) -
\delta\varphi(1)$, where 1 denotes a point at the last scattering
surface in the decoupling epoch and 2 observation point.
$\delta\varphi(2)$ is the variation/fluctuation at the last
scattering surface. $\delta\varphi(1)$ at the present observation
point is fixed. Therefore the covariance of fluctuation
$<[\delta\varphi(2) - \delta\varphi(1)]^2>$ gives the covariance of
$\delta\varphi^2(2)$ at the last scattering surface. Since our
Universe is isotropic to $\sim 10^{-5}$, this covariance is $\sim
(\xi \times 10^{-5})^2$ where the parameter $\xi$ depends on various
cosmological models \cite{ref10,ref17}.

In 2002, DASI microwave interferometer observed the polarization of
the cosmic background \cite{ref11}. E-mode polarization is detected
with 4.9 $\sigma$. The TE correlation of the temperature and E-mode
polarization is detected at 95$\%$ confidence. This correlation is
expected from the Raleigh scattering of radiation. However, with the
(pseudo)scalar-photon interaction (2), the polarization anisotropy
is shifted differently in different directions relative to the
temperature anisotropy due to propagation; the correlation will then
be downgraded. In 2003, from the first-year data (WMAP1), WMAP found
that the polarization and temperature are correlated to more than 10
$\sigma$ \cite{ref12}. This gives a constraint of about $10^{-1}$
for $\Delta\varphi$ \cite{ref9,ref18}.

Further results [13-16] and analyses [15, 19-27] of CMB polarization
observations came out after 2006.  In Table 1, we update our
previous compilations of \cite{ref10,ref17}. Although these results
look different at 1 $\sigma$ level, they are all consistent with
null detection and with one another at 2 $\sigma$ level. For the
interpretation of cosmic polarization rotation in various cosmologic
models, please see \cite{ref10,ref17}.

The Faraday rotation due to magnetic field is wavelength-dependent
while the cosmic polarization rotation due to effective
pseudoscalar-photon interaction is wavelength-independent. This
property can be used to separate the two effects in more precise
observations.

\begin{table}[ph]
\centerline{\begin{tabular}{|c|c|c|}
  \hline & & \\[-6pt]
 Reference & Constraint [mrad] & Source data  \\[3pt]
  \hline & & \\[-6pt]
  Ni \cite{ref9,ref18} & $\pm100$ & WMAP1 \cite{ref12} \\[3pt]
  \hline & & \\[-6pt]
   Feng, Li, Xia, Chen, and Zhang \cite{ref19} & $-$105 $\pm$ 70  & B03 \cite{ref14}  \\[3pt]
 \hline & & \\[-6pt]
  Liu, Lee, Ng [20]&$\pm $24 & B03 \cite{ref14}\\[3pt]
  \hline & & \\[-6pt]
  Kostelecky and Mews \cite{ref21}& 209 $\pm $ 122  & B03 \cite{ref14} \\[3pt]
  \hline & & \\[-6pt]
 Cabella, Natoli and Silk \cite{ref22}  & $-$43 $\pm $ 52  & WMAP3 \cite{ref13}\\[3pt]
  \hline & & \\[-6pt]
 Xia, Li, Wang, and Zhang \cite{ref23} &  $-$108 $\pm $ 67& WMAP3 \cite{ref13} $\&$ B03 \cite{ref14}\\[3pt]
   \hline & & \\[-6pt]
 Komatsu, $\emph{et al}$. \cite{ref15}& $-$30 $\pm $ 37& WMAP5 \cite{ref15} \\[3pt]
 \hline & & \\[-6pt]
 Xia, Li, Zhao, and Zhang \cite{ref24}& $-$45 $\pm $ 33  & WMAP5 \cite{ref15} $\&$ B03 \cite{ref14}\\[1pt]
 \hline & & \\[-6pt]
 Kostelecky and Mews \cite{ref25}& 40 $\pm $ 94 & WMAP5 \cite{ref15}\\[3pt]
 \hline & & \\[-6pt]
 Kahniashvili, Durrer, and Maravin  & $\pm $ 44 &  WMAP5 [15]\\[3pt]
 \hline & & \\[-6pt]
 Wu, $\emph{et al}$. \cite{ref27} & 9.6 $\pm $ 14.3 $\pm $ 8.7 & QuaD \cite{ref16}\\[3pt]
\hline
\end{tabular}}
 \caption{Constraints on cosmic polarization rotation from
CMB (cosmic microwave background).} \end{table}\label{table2}

\section{Discussion and Outlook}
Better accuracy in CMB polarization observation is expected from
PLANCK mission  launched on May 14, 2009. Dedicated CMB polarization
observers like B-Pol mission, CMBpol mission and LiteBIRD mission
would improve the sensitivity further. These development would probe
the fundamental issues of effective pseudoscalar-photon interaction
discussed in this paper more deeply in the future. \vspace{5 mm}

 We would like to thank the National Science Council (Grant Nos
NSC97-2112-M-007-002 and NSC98-2112-M-007-009) and the National
Natural Science Foundation (Grant Nos. 10778710 and 10875171) for
supporting this work.


\begin{footnotesize}



%

\end{footnotesize}


\end{document}